\documentstyle[epsfig,12pt,a4p,subfigure]{article}

\newcommand{\pipipipi}{\mbox{$\pi^+\pi^-\pi^+\pi^-$ }}

\newcommand{\pipi}{\mbox{$\pi^{+}\pi^{-}$} }

\newcommand{\fmeson}{\mbox{$f_{2}(1270)$} }

\begin{document}
\begin{titlepage}
\def\footnoterule{\hrule width 1.0\columnwidth}
\begin{center} {\large EUROPEAN ORGANIZATION FOR NUCLEAR RESEARCH}
 \end{center}
\begin{tabbing}
put this on the right hand corner using tabbing so it looks
 and neat and in \= \kill
\> {ALICE/98-45}   \\
\> {3 November 1998}
\end{tabbing}
\bigskip
\bigskip
\begin{center}{\Large  {\bf
A study of Double Pomeron Exchange in ALICE}
}\end{center}
\bigskip
\bigskip
\begin{center}{
A.\thinspace Kirk \\
and \\
O. Villalobos Baillie
}\end{center}
\bigskip
\bigskip
\begin{tabbing}
aba \=   \kill
\> \small
School of Physics and Astronomy, University of Birmingham, Birmingham, U.K. \\
\end{tabbing}
\begin{center}{\bf {{\bf Abstract}}}\end{center}

{
The non-Abelian nature of QCD suggests that
particles that have a gluon constituent, such as glueballs or
hybrids, should exist.
Experiments WA76, WA91 and WA102
have performed a dedicated search for these states
in central production using
the CERN Omega Spectrometer.
New results from central production show that
there is a kinematical filter which can select
out glueball candidates from known $q \overline q$  states.
A further study of this at high energies
is essential in order to get information on
the $M(X^0)$~$>$ 2 GeV region.
This paper describes how this could be done using the
the ALICE detector at the LHC.
}
\bigskip
\bigskip
\end{titlepage}
\setcounter{page}{2}

\section {Introduction}
Recent developments in the study of hadronic interactions show that
central production is a mechanism which can be used to great advantage
in the study of hadronic spectra. The current studies have been
performed in fixed target experiment at $\sqrt s \sim$ 20 GeV.  It
would be of great interest to extend these studies to higher energies,
where it should be much easier to disentangle the production mechanism.
This paper is organized as follows.  In section 2 the spectrum of
non-$q \overline{q}$ mesons is discussed, in section 3 the results
from the Omega Central Production experiments are reviewed, and in
section 4 a possible extension of these studies using the ALICE
detector in pp mode is presented.
\section{The Glueball Spectrum}
\par
The present understanding of strong interactions is that they are
described
by Quantum ChromoDynamics (QCD). This non-Abelian field theory
not only describes how quarks and antiquarks interact, but also
predicts that the gluons which are the quanta of the field will themselves
interact to form mesons.
If the object formed is composed entirely of valence gluons the meson
is called a glueball, however if it is composed of a mixture of
valence quarks, antiquarks
and gluons (i.e. $ q \overline q g$ ) it is called a hybrid.
In addition, $ q \overline q q \overline q $ states are also predicted.
\par
The best estimate for the masses of glueballs comes from
lattice gauge theory calculations
\cite{re:lgt}
which show that the lightest glueball has $J^{PC}$~=~$0^{++}$ and that
\begin{center}
$m(2^{++})/m(0^{++}) =1.5 $
\end{center}
and  depending on the extrapolation used from the lattice parameters to
mass scale that
\begin{center}
$m(0^{++}) =(1500-1750) $MeV.
\end{center}
The mass of the $0^{-+}$ glueball is predicted to be similar to that of the
$2^{++}$ glueball whilst glueballs with other quantum numbers
are predicted to be higher in mass.
\par
The flux tube model has been used to calculate the masses of the lowest
lying hybrid states and recent predictions
\cite{re:ISGURBNL}
are that
\begin{center}
$m(1^{--},0^{-+},1^{-+},2^{-+}) \approx 1900$~MeV.
\end{center}
\par
Hence, these non-$q \overline q$ states
are predicted to be in the same mass range
as the normal  $q \overline q$ nonet members
and hence we need a method of identifying them.
\par
The following have been suggested as possible ways to identify
gluonic states.
\begin{itemize}
\item
To look for "oddballs": States with $J^{PC}$ quantum numbers not allowed
for normal $ q \overline q$ states. For example $J^{PC} = 1^{-+}$.
\item
However the lightest non-$q \overline q$ states
are predicted to have the same quantum numbers as $q \overline q$ states.
Therefore we need
to look for extra states, that is states that have quantum numbers
of already completed nonets and that have masses which are sufficiently
low that they are unlikely to be members of the radially excited nonets and
hence they can not be described as being pure $ q \overline q $ states.
\item
If extra states are found then in order to isolate which state is the
likely non-$q \overline q$ state we can
\begin{enumerate}
\item[a)]
Look for states with unusual branching ratios.
\item[b)]
Look for states preferentially produced in gluon rich processes.
These processes are described below.
\end{enumerate}
\end{itemize}
\par
\begin{figure}[tp]
\begin{center}
\epsfig{file=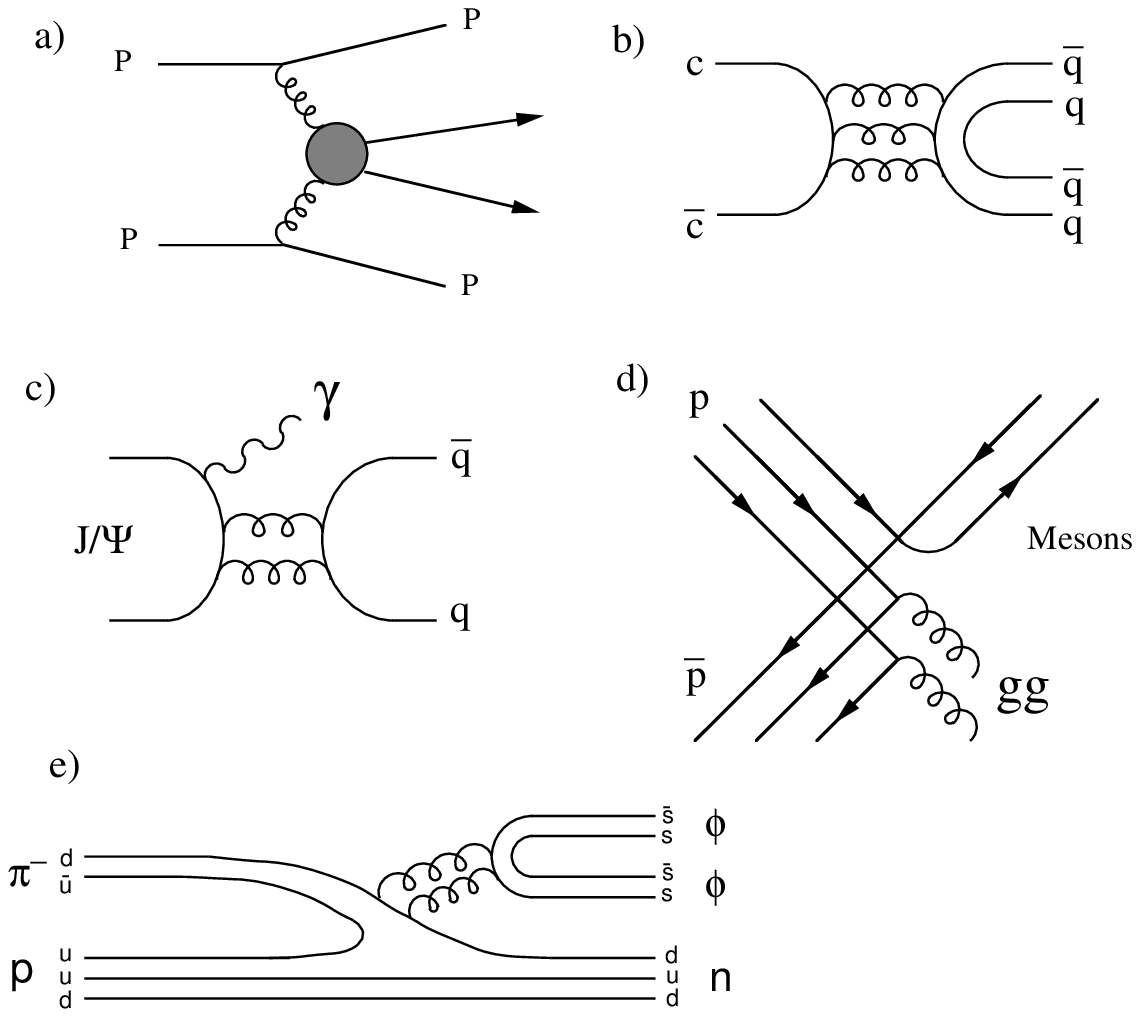,height=16cm,width=16cm}
\caption{Gluon rich channels.
Dynamical configurations that have been used to study light hadron spectroscopy
in a search for glueball states.}
\label{fi:dysum}
\end{center}
\end{figure}

Fig.~\ref{fi:dysum}
summarises several dynamical configurations which have been suggested
as possible
sources of gluonium and where experiments have been performed.
\begin{enumerate}
\item
Pomeron-Pomeron scattering is shown in fig.~\ref{fi:dysum}a).
The Pomeron is
an object which can be described as
a multi-gluon state, and
is thought to be
responsible for the large cross sections of diffractive reactions.
Consequently
Double Pomeron Exchange (DPE) is considered to be
a possible source of glueballs.
\item
The $J/\psi$ decay is believed to be
a highly glue rich channel either via the hadronic decay
shown in fig.~\ref{fi:dysum}b), or via the radiative decay shown in
fig.~\ref{fi:dysum}c).
\item
Figure~\ref{fi:dysum}d)
shows proton-antiproton annihilation; the annihilation region
of quarks and antiquarks is a source of gluons where glueballs and hybrids
could be produced.
\item
Special hadronic reactions, an example of which is shown in
fig.~\ref{fi:dysum}e)
where the $\phi\phi$ system is thought to be
produced via an intermediate state containing
gluons. Reactions of this kind which have disconnected quark lines
are said to be OZI violating
\cite{re:OZI}.
\end{enumerate}
\par
The first reaction is the one studied
by experiments WA76, WA91 and WA102 at the Omega spectrometer.
In the following section the  status of these experiments is
reviewed and the possibility
of a glueball-$q \overline q$ filter in central production is discussed.
\section{The Omega Central production experiments}
\subsection{Introduction}
\par
The Omega central production experiments
(WA76, WA91 and WA102) are
designed to study exclusive final states
formed in the reaction
\begin{center}
pp$\longrightarrow$p$_{f}X^{0}$p$_s$,
\end{center}
where the subscripts $f$ and $s$ refer to the fastest and slowest
particles in the laboratory frame respectively and $X^0$ represents
the central system. Such reactions are expected to
be mediated by double exchange processes
where both Pomeron and Reggeon exchange can occur.
\par
The trigger was designed to enhance double exchange
processes with respect to single exchange and elastic processes.
Details of the trigger conditions, the data
processing and event selection
have been given in previous publications~\cite{re:expt}.
\subsection{The possibility of a Glueball-$q \overline q$ filter in central
production }
\par
The experiments have been
performed at incident beam momenta of 85, 300 and 450 GeV/c, corresponding to
centre-of-mass energies of
$\sqrt{s} = 12.7$, 23.8 and 28~GeV.
Theoretical
predictions \cite{pred} of the evolution of
the different exchange mechanisms with centre
of mass energy, $\sqrt{s}$, suggest the following behaviour for the
cross sections:
\begin{center}
$\sigma$(RR) $\sim s^{-1}$,\\
$\sigma$(RP) $\sim s^{-0.5}$,\\
$\sigma$(PP) $\sim$ constant,
\end{center}
where RR, RP and PP refer to Reggeon-Reggeon, Reggeon-Pomeron and
Pomeron-Pomeron
exchange respectively. Hence we expect Double Pomeron Exchange
(DPE) to be more significant at high energies, whereas the Reggeon-Reggeon and
Reggeon-Pomeron mechanisms will be of decreasing importance.
The decrease of the non-DPE cross section with energy can be inferred
from data
taken by the WA76 collaboration using pp interactions at $\sqrt{s}$ of 12.7 GeV
and 23.8 GeV \cite{wa76}.
The \pipi mass spectra for the two cases show that
the signal-to-background ratio for the $\rho^0$(770)
is much lower at high energy, and the WA76 collaboration report
that the ratio of the $\rho^0$(770) cross sections at 23.8 GeV and 12.7 GeV
is 0.44~$\pm$~0.07.
Since isospin 1 states such as the $\rho^0$(770) cannot be produced by DPE,
the decrease
of the $\rho^{0}(770)$ signal at high $\sqrt{s}$
is consistent with DPE becoming
relatively more important with increasing energy with respect to other
exchange processes.
\par
\begin{figure}[htp]
\begin{center}
\mbox{\begin{tabular}{l}
\subfigure{\mbox{\epsfig{file=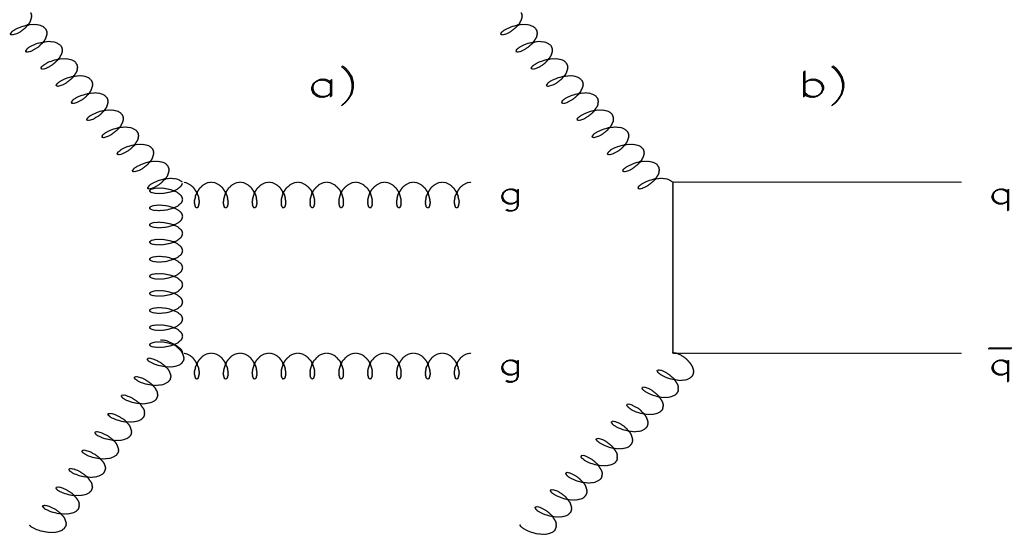,height=6cm,
width=14.0cm,bbllx=0pt,bblly=0pt,bburx=550pt,bbury=300pt}}}  \\
\end{tabular}}
\end{center}
\caption{Schematic diagrams
of the coupling of the exchange particles into the final state meson
for a) gluon exchange and b) quark exchange.}
\label{fi:feyn1}
\end{figure}

However,
even in the case of pure DPE
the exchanged particles still have to couple to a final state meson.
The coupling of the two exchanged particles can either be by gluon exchange
or quark exchange. Assuming the Pomeron
is a colour singlet gluonic system if
a gluon is exchanged then a gluonic state is produced, whereas if a
quark is exchanged then a $q \overline q $ state is produced
(see figures~\ref{fi:feyn1}a) and b) respectively).
It has been suggested recently~\cite{closeak} that
for small differences in transverse momentum between the two
exchanged particles
an enhancement in the production of glueballs
relative to $q \overline q$ states may occur.
The difference in the transverse momentum vectors ($dP_T$) is defined to be
\begin{center}
$dP_T$ = $\sqrt{(P_{y1} - P_{y2})^2 + (P_{z1} - P_{z2})^2}$
\end{center}
where
$Py_i$, $Pz_i$ are the y and z components of the momentum
of the $ith$ exchanged particle in the pp centre of mass system.
\par
\begin{figure}[ht]
\begin{center}
\epsfig{file=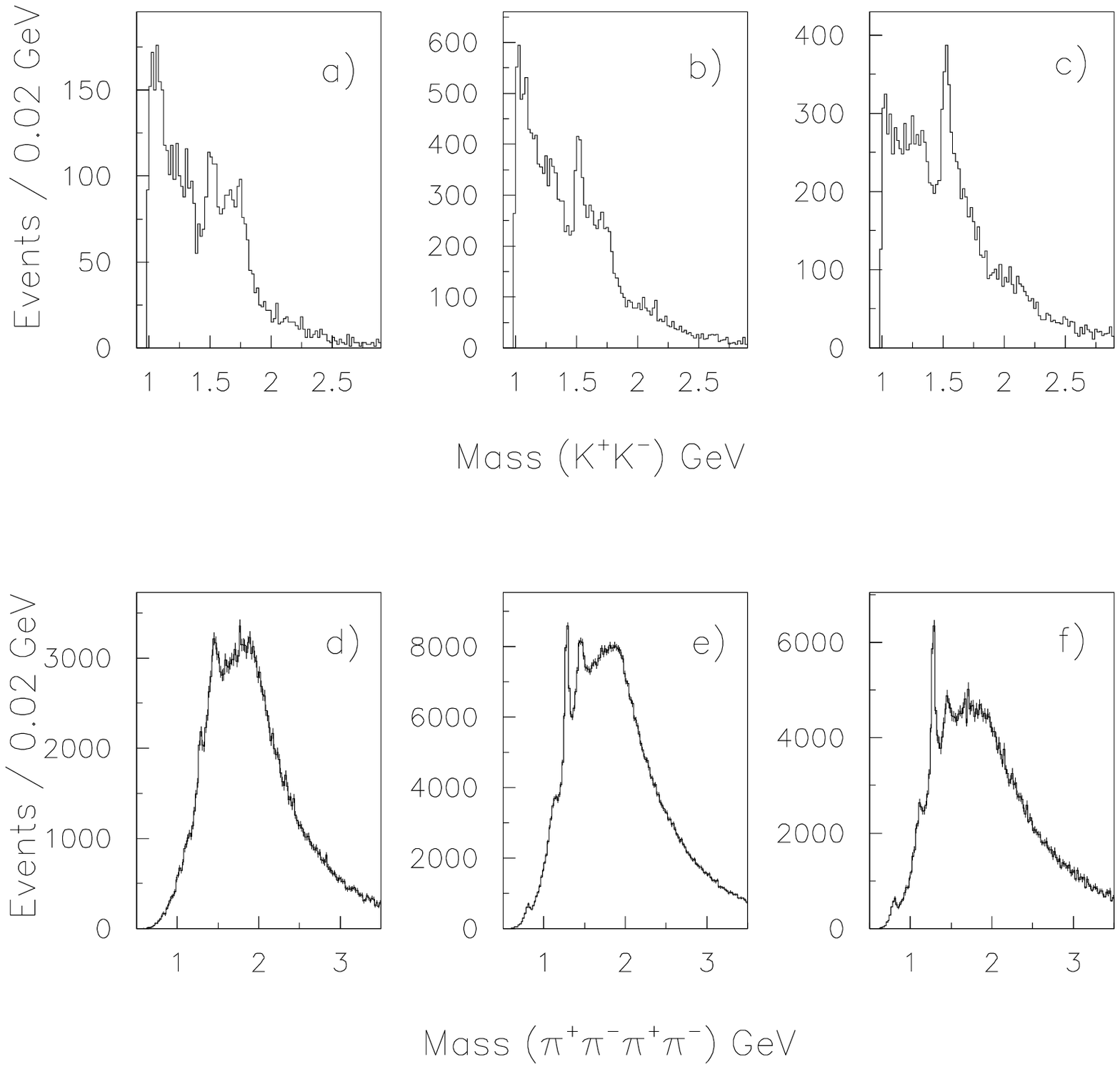,height=22cm,width=17cm}
\end{center}
\caption{$K^+K^-$ mass spectrum for a) $dP_T <   0.2$ GeV,
b) $0.2 <   dP_T <   0.5$ GeV and c) $dP_T >   0.5$ GeV and
the \pipipipi mass spectrum for d) $dP_T <   0.2$ GeV,
e) $0.2 <   dP_T <   0.5$ GeV and f) $dP_T >   0.5$ GeV.}
\label{fi:2k}
\end{figure}

Figures~\ref{fi:2k}a), b) and c) show the effect of the
$dP_T$ cut on the $K^+ K^-$ mass spectrum where
structures can be observed in the 1.5 and 1.7 GeV mass region which have
been previously identified as the
$f_{2}^\prime$(1525) and the $f_J(1710)$~\cite{re:WA76KK}.
As can be seen,
the $f_{2}^\prime$(1525) is produced dominantly at high $dP_T$,
whereas the $f_J(1710)$ is produced dominantly at low $dP_T$.
\par
In the \pipipipi  mass spectrum a dramatic effect is observed,
see figures~\ref{fi:2k}d), e) and f).
The $f_1(1285)$ signal has virtually disappeared at low $dP_T$
whereas
the $f_0(1500)$ and $f_2(1900)$ signals remain.
Similar effects are observed in all the other channels analysed to
date~\cite{NEWKKPI,NEW3PI}.
In fact it has been observed that
all the undisputed
$ q \overline q $ states
(i.e. $\rho^0(770)$, $\eta^{\prime}$, \fmeson, $f_1(1285)$,
$f_2^\prime(1525)$ etc.)
are suppressed as $dP_T$ goes to zero,
whereas the glueball candidates
$f_J(1710)$, $f_0(1500)$ and $f_2(1900)$ survive.
It is also interesting to note that the
enigmatic
$f_0(980)$,
a possible non-$q \overline q$ meson or $K \overline K$ molecule, does not
behave as a normal $q \overline q$ state.
\par
A Monte Carlo simulation of the trigger, detector acceptances
and reconstruction program
shows that there is very little difference in the acceptance as a function of
$dP_T$ in the different mass intervals considered
within a given channel and hence the
observed differences in resonance production can not be explained
as acceptance effects.
\par
\subsection{Summary of the effects of the $dP_T$ filter}
\par
\begin{figure}[htp]
\begin{center}
\epsfig{file=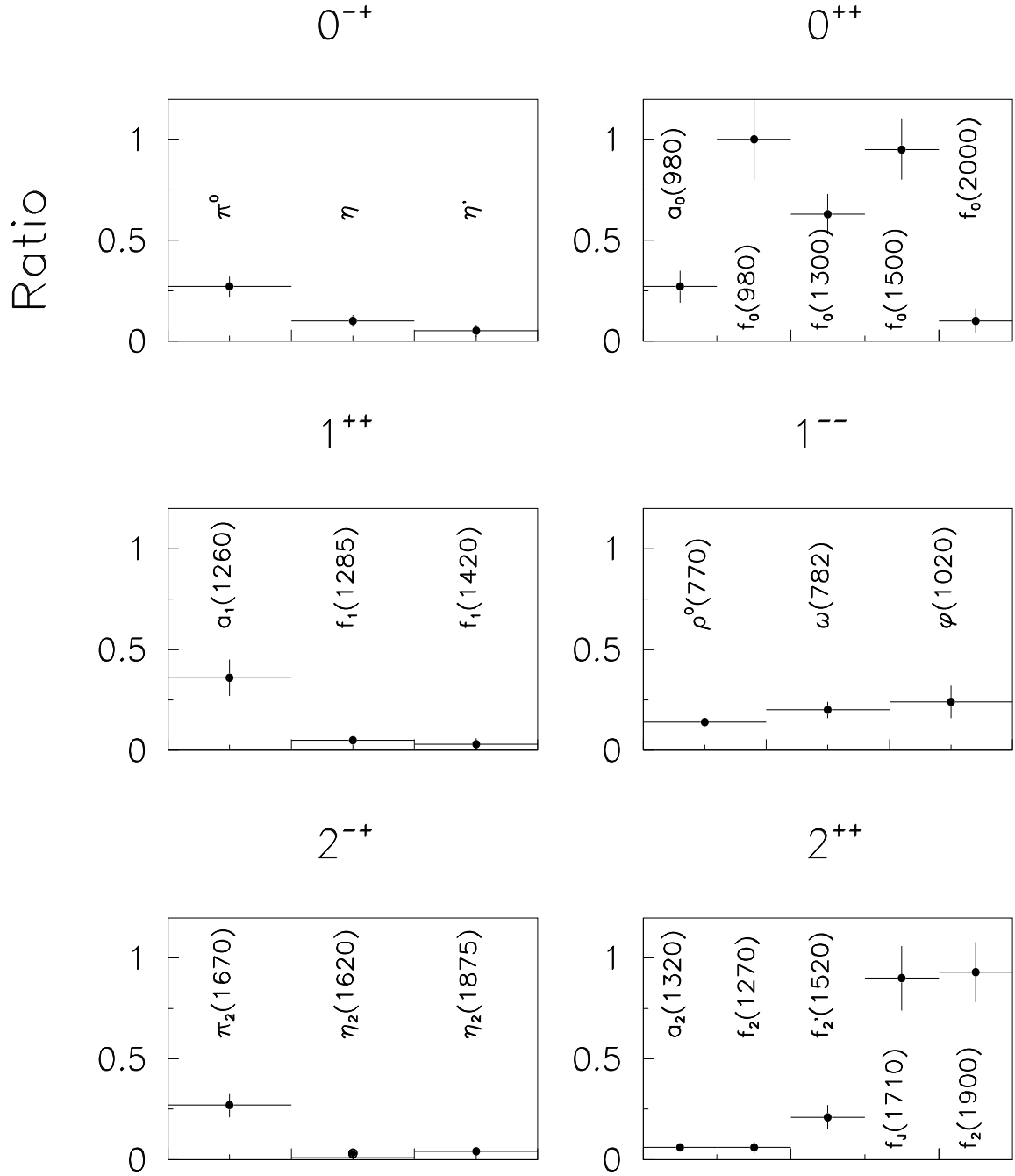,height=22cm,width=16cm}
\end{center}
\caption{The ratio of the amount of resonance with
$dP_T$~$\leq$~0.2 to the amount with
$dP_T$~$\geq$~0.5~GeV.
}
\label{fracratio}
\end{figure}

The contribution of each resonance as a function
of $dP_T$  has been calculated.
Figure~\ref{fracratio} shows the ratio of the number of events
for $dP_T$ $<$ 0.2 GeV (the glue rich exchange region) to
the number of events
for $dP_T$ $>$ 0.5 GeV (the quark rich exchange region)
for each resonance considered.
It can be observed that all the undisputed $q \overline q$ states
which can be produced in DPE, namely those with positive G parity and $I=0$,
have a very small value for this ratio ($\leq 0.1$).
Some of the states with $I=1$ or G parity negative,
which can not be produced by DPE,
have a slightly higher value ($\approx 0.25$).
However, all of these states are suppressed relative to the
interesting states, i.e. those which could have a gluonic component, which have
a large value for this ratio.

\subsection{Conclusions}
\par
The results observed in central production to date indicate
the possibility of a
glueball-$q \overline q$ filter mechanism in central production.
All the
undisputed $q \overline q $ states are observed to be suppressed
at small $dP_T$, but the glueball candidates
$f_0(1500)$, $f_J(1710)$, and $f_2(1900)$ ,
together with the enigmatic $f_0(980)$,
survive.
\section{Possible studies in ALICE}
\subsection{Introduction}
\par
A new effect has been observed in pp collisions and deserves
to be studied more fully. Performing these studies
at LHC energies has two very useful consequences.
Firstly the mass range of the centrally produced system
is given by
\begin{center}
$M^2$~=~$s(1-x_{F1})(1-x_{F2})$
\end{center}
hence by increasing s the mass range (M) is increased.
The second feature is that the exchange will be effectively
100 \% double Pomeron i.e. there will no contamination
from Reggeon exchange. This second feature should help greatly
in understanding the underlying dynamics.
\subsection{What can be done in ALICE}
\par
Due to the limited $\gamma$ detection in ALICE only all
charged decays of the centrally produced system can be studied namely:
$\pi^+ \pi^-$, $K^+K^-$, $\pi^+\pi^-\pi^+\pi^-$,
$K^+K^-\pi^+\pi^-$, $K^+K^-K^+K^-$ and $e^+e^-$.
Although this is a limited number of channels
it is interesting to note that all the glueball candidates
observed to date have been seen in one or more of these decay
modes.
\par
One of the major requirements of this study is that we are
able to reconstruct exclusive events i.e we can exclude decays
that have $\pi^0$s in their final state by using momentum balance.
This of course requires that we observe all the charged tracks
including the outgoing protons.
\par
In order to exclude decays that involve a $\pi^0$ we need to require
that the missing $P_T$~$\le$~200~MeV. In order to do this
we need
\begin{itemize}
\item the spread in the transverse momentum of the incident beam momentum
be small and
\item small measurement errors on the outgoing protons.
\end{itemize}
\subsubsection{Incident Beam}
\par
The spread in the momentum of the incident beam
is determined by the $\beta^*$ of the final focus.
For a $\beta^*$ of 0.5 m the spread in the transverse momentum ($\delta P_T$)
of the incident beam is
$\delta P_T$~=225~MeV.
Hence a $\beta^*$ of 0.5 m would be unacceptable.
For
a $\beta^*$ of 250 m the
$\delta P_T$ is 10~MeV which would be acceptable.
This is in agreement with the plans to
run ALICE with the maximum
possible $\beta^*$ during pp running in order to
reduce the luminosity.
\subsubsection{Outgoing protons}
\par
The two outgoing protons could be measured using two stations of
Roman pots. The exact position of these stations need to be
determined. Initial simulations show that we
can accept protons that are scattered between 30 and 180~$\mu$rad.
Where the inner limit is defined by the
10$\sigma$ profile limit of the beam and the outer limit
by the beam pipe.
Assuming that each station was composed of several planes
of 10$\mu$m pitch microstrip detectors
then the outgoing proton could be measured with a
precision of
\begin{center}
$\delta P_T$~=~50~MeV and $\delta P_L$~=~7~GeV.
\end{center}
These values would be sufficient for our requirements.
\subsection{Machine optics}
\par
\begin{figure}[htp]
\begin{center}
\epsfig{file=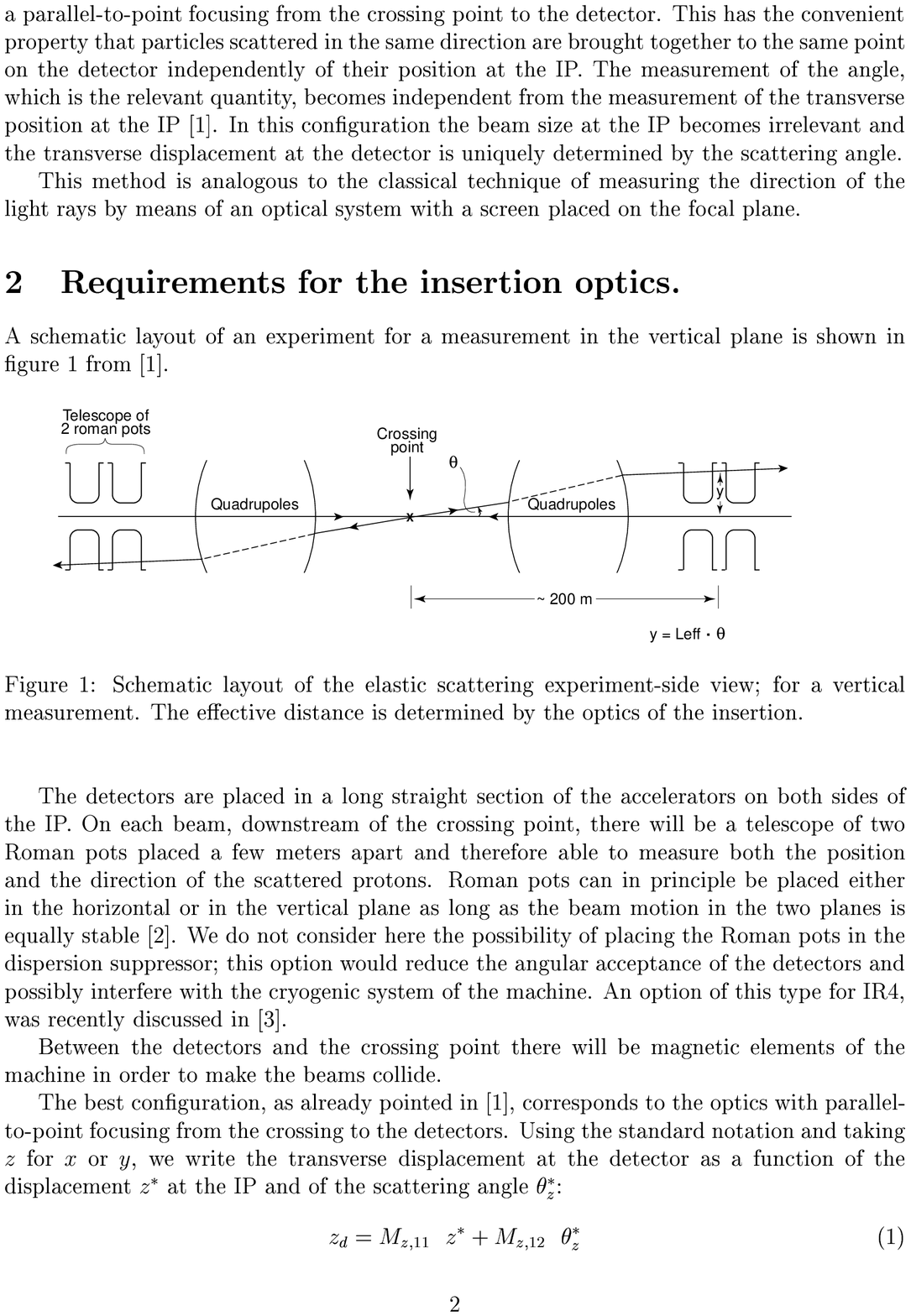,height=6cm,width=14.0cm,
bbllx=50pt,bblly=405pt,bburx=550pt,bbury=540pt,clip=}
\end{center}
\caption{Schematic layout of the Roman pots.
}
\label{fi:interxs}
\end{figure}

The requirements of the machine optics have been studied by the
TOTEM collaboration~\cite{TOTEM,LHC1,LHC2} who have
a similar requirement of measuring the scattered protons.
A schematic layout for the Roman Pot detectors is shown in
fig.~\ref{fi:interxs}. The detectors are placed in a long
straight section of the accelerator on both sides of the intersection point.
On each beam there will be a telescope of two Roman Pots placed 130 meters
apart and therefore able to measure both the position and
direction of the scattered protons.
Between the detectors and the crossing point there will be magnetic elements
of the machine which are used to make the machine collide.
In order to get the best measurement of the scattered protons
the best configuration of these magnetic elements
is the so called "parallel-to-point" focusing configuration.
\par
The minimum distance of approach of the inner edge of the detectors to the
beam axis, $z_d$, is proportional to the r.m.s. beam size at the
detector position, $\sigma_{z_d}$ i.e. $z_d = k \sigma_{z_d}$.
At the SPS collider it was found empirically by the UA4 experiment that
$k \approx 15 - 20$.
\par
For protons which hit the detector on the inner edge, the scattering angle
will be
\begin{equation}
\theta_{z_d} = k \sqrt \frac{\epsilon_z}{\beta^*_z}
\end{equation}
where $\epsilon_z$ is the emittance which has a nominal value of
$5.0 \times 10^{-10}$ m rad~\cite{LHC3}. The useful minimum angle is
given by $\theta_{z_{min}}= \sqrt 2 \theta_{z_d}$.
For a $\beta^*_z=250m$, $\theta_{z_{min}}=30 \mu$rad.
\par
An additional requirement comes from the fact that the actual distance $z_d$
of the inner edge of the detector from the machine axis should not be too
small, in order to avoid problems from possible beam instabilities.
Assuming a minimum acceptable value for $z_d$ of 1.5 mm~\cite{TOTEM}
then
\begin{equation}
L_{z_{eff}}\theta_{z_d} \ge z_{d_{min}} = 1.5 {\rm mm}
\end{equation}
which fixes the effective distance of the detectors to be
$L_{z_{eff}} \ge 50 $m.
\par
\begin{figure}[htp]
\begin{center}
\epsfig{file=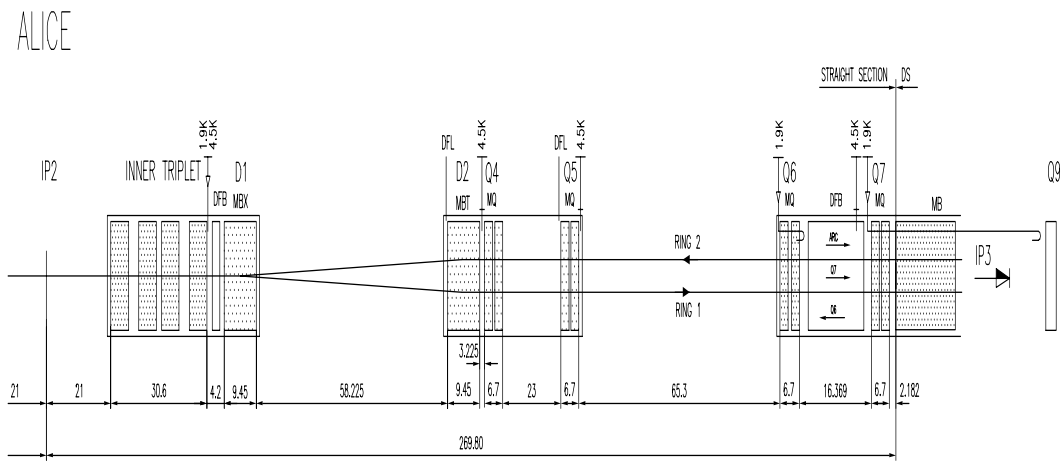,height=14cm,width=6cm,angle=90}
\end{center}
\caption{Layout of the right part of intersection region 2.
}
\label{fi:inter2}
\end{figure}

The layout of the right hand side of intersection point 2
is shown in fig.~\ref{fi:inter2}. There are in principle two possible
locations for the detectors, one just before dipole D2 and the second
between Q5 and Q6.
The position at D2 has the advantage that the tracking of the scattered protons
is easier because they transverse only four magnets.
The advantage of the Q5-Q6 setup is a larger lever arm between the
two detectors in the Roman pot telescope.
\subsection{Simulations}
The reaction
\begin{center}
pp$\longrightarrow$p$X^{0}$p,
\end{center}
at $\sqrt s$~=~14~TeV,
has been generated using a modified version of the WA102
event generator.
The $x_F$ distribution of the exchange particles
has been assumed to scale as $1/\sqrt s$ and the
$t$ slope of the proton vertex, which is parameterised
as $e^{-bt}$, with  $b$~=~$b_0ln(s/s_0)$ where $b_0$~=~6~GeV$^{-2}$
and $s_0$~=~784~GeV$^2$. The four momentum transfer distribution
from the proton vertex then has a distribution of the form
$e^{-24t}$.
\par
The outgoing protons
are required to be detected in Roman pot detectors and the
centrally produced system, $X^0$, is decayed into 2 or 4 particles
which are required to be within the acceptance of the central
tracking detectors.
\subsection{Detector and trigger requirements}
\par
The cross section for DPE at $\sqrt s$~=~30~GeV is
$\sigma(PP)$~=~140~$\mu$b.
Assuming the cross section scales as $s^{0.08}$ then the cross section
at $\sqrt s$~=~14~TeV is $\sigma(PP)$~=~370~$\mu$b.
In order to trigger on the centrally produced events we would need
a low multiplicity trigger in the central region to allow
us to detect events composed of two or four tracks.
An additional trigger would require two outgoing protons
that had scattered into the Roman pot detectors.
\par
\begin{figure}[htp]
\begin{center}
\epsfig{file=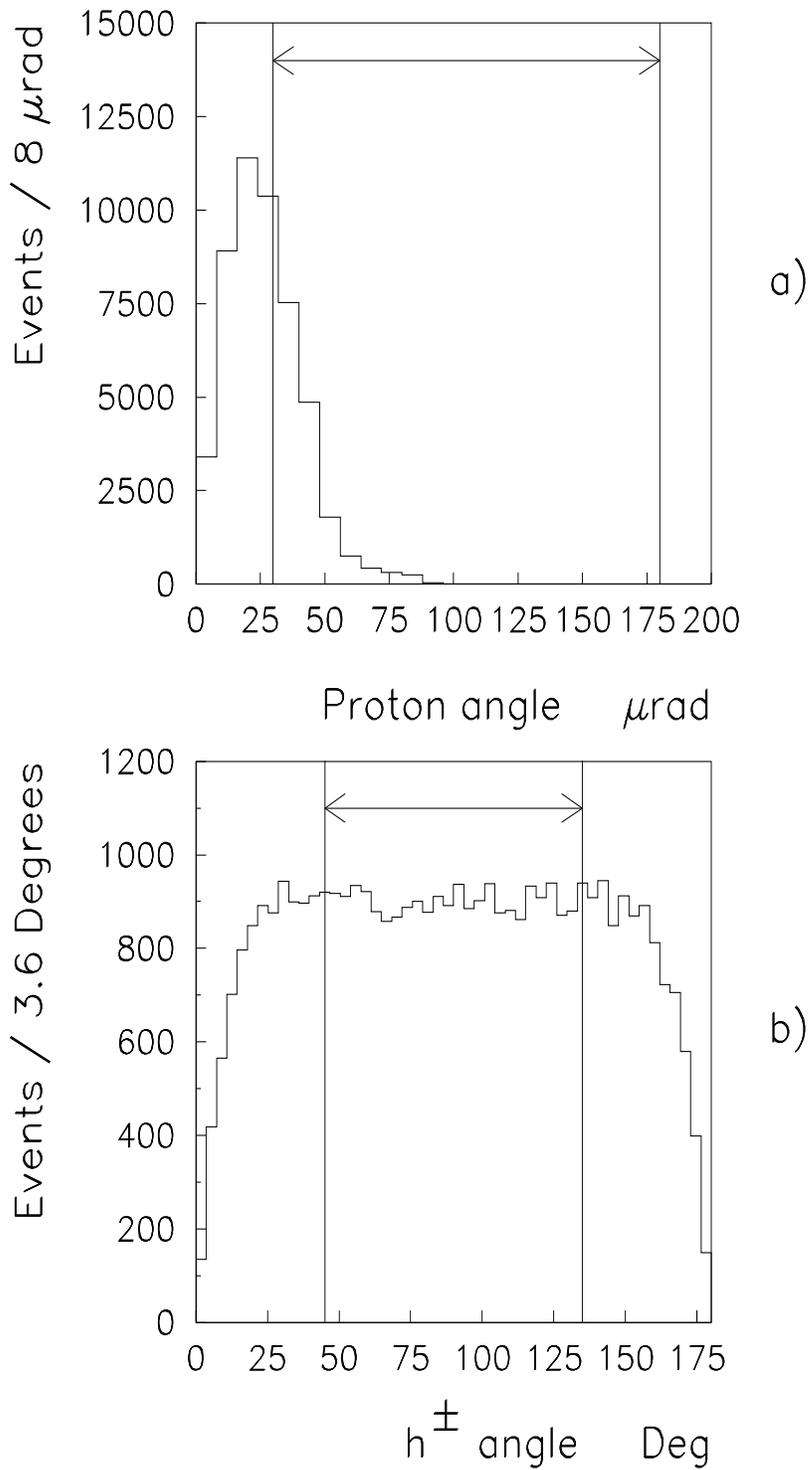,height=22cm,width=16cm}
\end{center}
\caption{Results of the simulation. a) The scattered angle of the protons
and b) the central system.
}
\label{fi:simul}
\end{figure}

Fig.~\ref{fi:simul}a) shows the distribution of the azimuthal
angle of the protons scattered in DPE collisions
together with the acceptance region of the Roman Pots.
Fig.~\ref{fi:simul}b) shows the distribution of the centrally produced
charged particle together with the acceptance of the central trackers.
This results in an acceptance for centrally produced events in the
2~GeV mass region of $\approx$~12.5~$\%$.
\par
Assuming a luminosity of $10^{30}$ cm$^{-2}$s$^{-1}$
we would have a trigger rate of 8~Hz from DPE events. This would
give us an integrated annual data sample (10$^7$ s) of 80 million events.

\subsection{The Central multiplicity trigger}
\par
A capability to trigger on one of the central detectors, for example the
pixel layers, seems advisable in order to select minimum bias p-p events.
We know that the mean multiplicity will be low.  At present triggering is
done using the FMD detectors only. These may have trouble rejecting
background if used
alone.  Furthermore, a realistic assessment of the noise requires
a very detailed Monte Carlo study, which has not yet been done and would
be very time consuming.
\par
A central trigger, particularly one looking for spatial correlations between
adjacent layers, complements the information from the FMD (time correlations
between different disks) very well, and should lead to efficient
background rejection.
\par
The practical problems concerning the implementation of such a trigger have
been discussed with the ITS group (Fabio Formenti).  The pixel chips can
have a trigger output, giving a response in ~50ns, and there are no
overwhelming obstacles to having such a trigger. However, having one clearly
makes the job of setting up the pixel layers considerably more difficult,
particularly from the point of view of accommodating the large numbers of
trigger cables coming out of the ITS area.  This is because the trigger
signals are not suitable for multiplexing.  For this reason, a solution in
which several pixel chips within a ladder are grouped to make a logical unit
with a larger area is the recommended solution.  A moderate amount of
Monte Carlo work should suffice to optimize the size of the logical units.
A consequence of this is that if this study
implies that the pixel trigger output is required, the ITS group should
be requested to include it in their TDR.
\par
The requirements for a Double Pomeron trigger would also be very well served
by such a pixel trigger.  Here one would require no activity in the FMD, but
expect to find the correlated hits in the forward particle stations at
level 2.

\subsection{Roman Pots for proton detection}
\par
Contact has been made with the TOTEM group
who have a similar Roman pot requirement.
They are well advanced with their
design and currently in favour of using the CMS detector.
They have performed
simulations of where to put the Roman pots for all four intersection
regions~\cite{LHC2,LHC3}.
In particular they have found that a modification to the
beam optics is required.
In the present design
the quadrapoles nearest to ALICE
are composed of
4 linked magnets working from one power
supply. In the modified design
only the middle two magnets are linked and
hence three power supplies are needed. This greatly helps
in measuring deflected protons. Of course such changes
need to be notified to the machine group as soon as possible.
\par
If TOTEM does go to CMS then we could profit from their developments of the
Roman pots and detectors. A very preliminary costing for the Roman
pots plus microstrip detectors would be 500 kChf.
\subsection{Additional advantage of having Roman Pot detectors}
\par
The addition of Roman Pot detectors to the ALICE setup would allow
an accurate measurement of the luminosity during p-p running in ALICE
which would enable better comparisons to be made with the ion-ion running.
\subsection{Summary}
\par
New results from central production show that
there is a kinematical filter which can select
out glueball candidates from known $q \overline q$  states.
A further study of this at high energies
is essential in order to get information on
the $M(X^0)$~$>$ 2 GeV region.
This could be done using the ALICE detector
by studying all charged decay modes.
It should be recalled that
all the known glueball candidates have been
observed in these decay modes.
\par
To perform this study we would require the addition of
a low multiplicity central
particle trigger, which is also needed to select minimum bias pp events.
We also need
the addition of Roman Pots which could also be used to give
an accurate measure of the luminosity in the ALICE detector.
It would allow a new area of physics to be explored
during proton-proton data taking.

\newpage
\begin{center}
{\bf Acknowledgements}
\end{center}
\par
We would like to thank A. Faus-Golfe, W. Kienzle and M. Haguenauer
for useful discussions.

\newpage
\newpage
\newpage
\clearpage
\newpage
\newpage
\clearpage

\end{document}